\documentclass[aps,prl,twocolumn,superscriptaddress]{revtex4-2}
\usepackage{amsmath}
\usepackage{amssymb}
\usepackage{graphicx}

\begin{document}

\title{Photon-Mediated Atomic Interactions in Curved Surface Settings}
%\title{Cooperative Resonances in Curved Surface Settings}


\author{Maor Carmi}
%\email{maor.carmi@mail.huji.ac.il}

\author{Michal Roth}
%\email{michal.roth@mail.huji.ac.il}

\author{Rivka Bekenstein}
\email{rivka.bekenstein@mail.huji.ac.il}

\affiliation{Racah Institute of Physics, Hebrew University of Jerusalem, Jerusalem 91904, Israel}

\begin{abstract}
Subwavelength atomic lattices have emerged as a promising platform for quantum applications, leveraging collective superradiant and subradiant effects to enhance light-matter interactions. Integrating atomic lattices into nanostructures is at the front of effort toward any application with atomic lattices, but is still a challenging theoretical task, as the dipole interactions are hard to model in arbitrary dielectric environment. Here, we develop a formalism that allows the embedding of emitters to a wide class of thin curved waveguides, by greatly simplifying the wave equation and consequentially the dipole interactions. We demonstrate the formalism for locally constant positive curved surfaces, where we derive a new surface Green function that allows exact computation of the collective superradiant and subradiant states. We also show how the curvature and the thickness of the waveguides affects the collective states through an effective surface wavelength.
\end{abstract}

%\keywords{}

\maketitle
%\section{\label{sec: intro}Introduction}

Cooperative effects in subwavelength atomic arrays have attracted considerable interest in the past years, as their extreme optical response makes them suitable for numerous applications in quantum technologies, such as stable quantum memories \cite{Kimble_2017}, atom-photon and atom-atom entanglement generation \cite{Bekenstein_Metasurface_2020, Nadav_2024, guimond2019subradiant}, platforms for quantum computation \cite{shah2024quantum} as well as enhanced coherent emission \cite{fernandez2022tunable, masson2022universality}. These cooperative effects originate from photon-mediated dipole-dipole interaction that are enhanced in these ordered atomic systems in free space \cite{Porras2008Aug,novotny2012principles}.
There is great interest in integration of atomic arrays and single atoms into nanostructures, such as nanocavities and waveguides \cite{nano_photonic_integration_0_Hannes_integrated_2024,nano_photonic_integration_1_goban2014atom,nano_photonic_integration_2_hood2016atom,nano_photonic_integration_3_thompson2013coupling,nano_photonic_integration_4_samutpraphoot2020strong}, as the atom-photon interaction strength is greatly enhanced in such settings \cite{nano_photonic_integration_4_samutpraphoot2020strong}, and it is considered a path toward scalable systems. 
However, theoretical investigations of subwavelength atomic arrays in dielectric structures is challenging, as the dyadic Green function, used to model the dipole-dipole interactions between the atomic emitters, has known analytic forms only for very specific cases \cite{green_1_multilayer_tomavs1995green,green_2_nanofiber_klimov2004spontaneous,meystre2007elements}, and one usually resorts to numerical methods. An approach for dealing with atomic arrays coupled to specific structured waveguides was successfully introduced in \cite{Ann_nanostructures_2017, tevcer2024strongly}, but arbitrary two-dimensional structures, which are required for practical and general nanophotonics, are still challenging to model analytically. 
Simultaneously, the study of optics on curved waveguides has gained significant interest mainly due to it being analogues to general relativity settings \cite{schultheiss2020light}. Specifically, it was shown that curved geometries allow for unique control over light propagation \cite{Batz_Peschel_2008, schultheiss2010optics} which has proved value for nanophotonics architectures \cite{Bekenstein_Curved_Space_2017,roth2024non}.
%, as well as quantum simulations \cite{kollar2019hyperbolic}. 

Here, we introduce a formalism that enables calculations of dipole-dipole interactions on two-dimensional waveguides, highlighting the role of the curvature of the surface. We do this by developing a curved-space formalism for finding the dyadic Green function that determines the dipole-dipole interactions in sub-wavelength atomic arrays. Specifically, we derive the wave equation for the magnetic vector potential constrained to an arbitrary surface, and show that for a wide class of surfaces the wave equation for the normal component can be decoupled, featuring an effective wavelength that is curvatures- and momentum-dependent. This greatly simplifies the wave equation, enabling the atomic array to be embedded in a wide range of two-dimensional geometries. To demonstrate our formalism we find a new closed-form dyadic Green function for locally spherical surfaces, and employ it for calculations of cooperative effects in curved settings. In addition, by analyzing a planar waveguide we show the effect of its geometry on the effective wavelength within the surface.\\
%\section{\label{sec: wave eqs} Vector Potential Wave Equation}
We consider an array of \(N\) closely spaced two-level emitters embedded in some arbitrary surface at positions \(\{\boldsymbol{r}_n\}_{n=1}^N\) . The cooperative effects in the array arise due to dipole-dipole interactions, which are most commonly described using the dyadic Green tensor of the electric field \cite{novotny2012principles}. Under the Markovian approximation, the system's dynamics are governed by the following effective non-Hermitian Hamiltonian \cite{Kimble_2017}:
\begin{equation}\label{eq: effective Hamiltonian}
\hat{H}_\text{eff}=\sum_{i=1}^N\hbar\omega\hat{\sigma}_{ee}^{\left(i\right)}-\mu_{0}\omega^{2}\sum_{i,j=1}^N\boldsymbol{\wp}_{i}^{\dagger}\boldsymbol{\overset{\leftrightarrow}{G}}\left(\boldsymbol{r}_{i},\boldsymbol{r}_{j}\right)\boldsymbol{\wp}_{j}\hat{\sigma}_{eg}^{\left(i\right)}\hat{\sigma}_{ge}^{\left(j\right)}
\end{equation}
where \(\sigma_{ab}^{\left(i\right)}\) are the atomic transition operators, \(\hbar\omega\) is the emitters' natural resonance frequency, \(\boldsymbol{\wp}_i\) are the vectorial dipole matrix elements of each emitter, \(\boldsymbol{\overset{\leftrightarrow}{G}}\) is the dyadic Green tensor and \(\mu_0\) is the permeability of free space. Diagonalizing the effective Hamiltonian gives the collective states of the array \cite{Kimble_2017}. The only part of the Hamiltonian that is geometry-dependent is the dyadic Green tensor \cite{Kimble_2017}, which is the main focus of this study. \\

\begin{figure}
\includegraphics[width=1\linewidth]{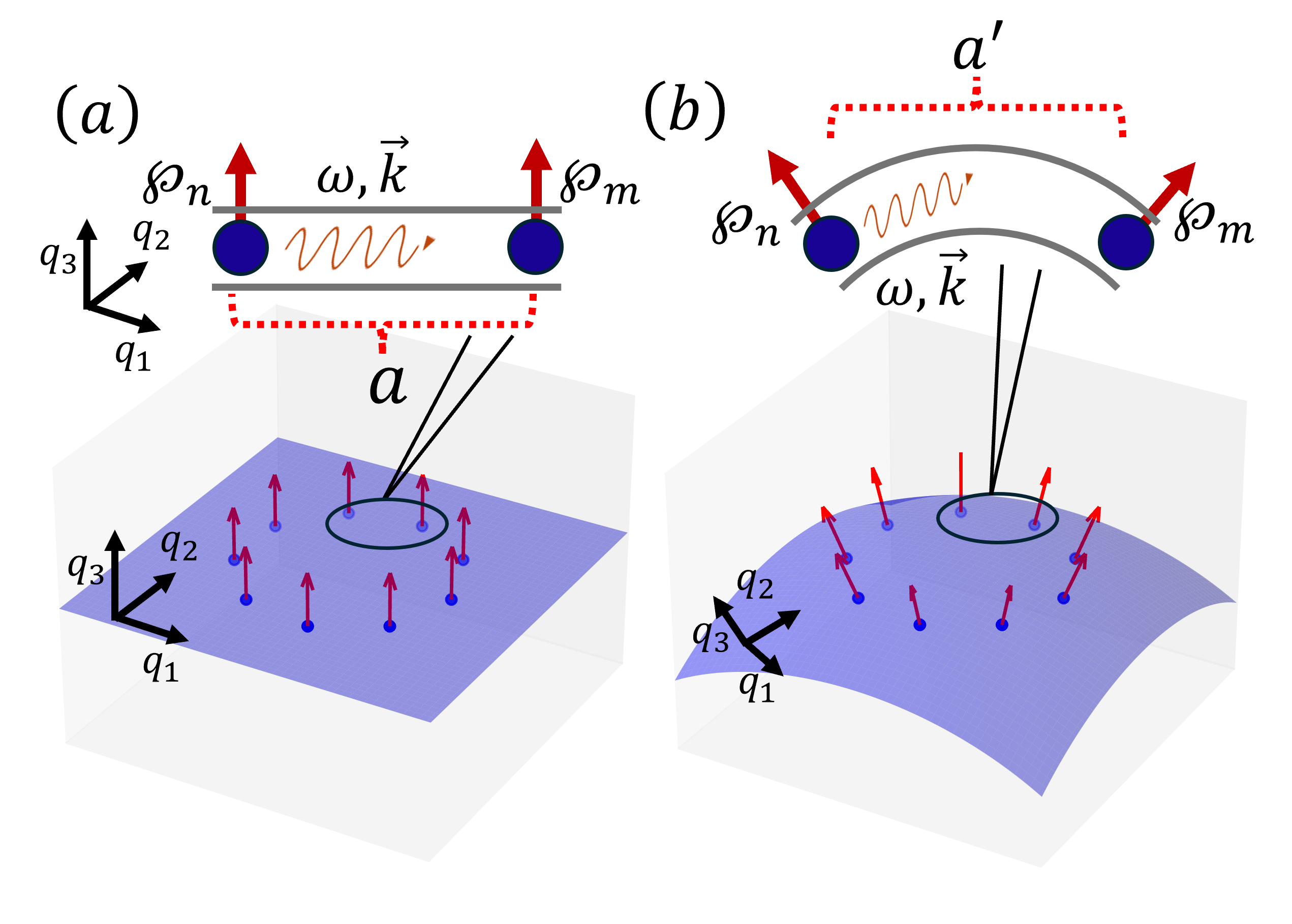}
\caption{\label{fig: schematics} Schematics of the systems under consideration. \(\hat{q}_a\) represent in-plane coordinate basis vectors (on the tangent plane), and \(\hat{q}_3=\hat{\boldsymbol{n}}\) is the unit normal vector. (a) ring of emitters constrained to a surface, with dipoles directed at the normal direction \(\hat{q}_3=\hat{z}\). The distance between each emitter is the lattice spacing \(a\), and the photon exchange interaction is shown to travel a straight path. (b) ring of emitters constrained to a spherical surface patch, with dipoles oriented to the normal direction, which is now radially outwards \(\hat{q}_3=\hat{r}\). Photon exchange is shown to travel a longer path, as now the distance calculated as the arc length across the sphere, where \(R\) is the sphere's radius and \(\alpha\) is the central angle between the emitters.}
\end{figure}

%We hereby derive the dyadic Green tensor for an arbitrary curved surface by deriving the wave equations for the magnetic vector potential. %As oppose to cartesian coordinates the equations for its different components are coupled, but for a wide class of surfaces the wave equation for the normal component can be decoupled from the tangential components.
The dyadic Green tensor is commonly derived using the potential fields \(A\) and \(\phi\) in the Lorentz gauge \cite{novotny2012principles}, which leads us to consider the vector-potential wave equation written in arbitrary space-coordinates (with Einstein's summation convention implied) \cite[p. 569]{Gravitation}:
\begin{equation}\label{eq: general wave equation}
   \left[\nabla^j\nabla_j+k^2\right]A^i+R^i_{\;j}A^j=-\mu_0 J^i
\end{equation}
where \(A^i\) are the components of the vector-potential, \(\nabla^j\nabla_j\)  is the vector Laplacian, \(k=\omega/c\) is the wavenumber, \( R^i_{\;j}\) is the Ricci curvature tensor and \(J^i\) is the three-current. The Ricci curvature tensor accounts for the curvature of space \cite{Gravitation}. In Cartesian coordinates (flat space), this equation reproduces the usual decoupled vector potential wave equation, as \(\nabla^j\nabla_j\) is reduced to a scalar Laplacian, and the Ricci tensor vanishes. However, we would like to express this equation in coordinates that naturally encode the structure of an arbitrary surface, which will most certainly be coupled.  The starting point of our derivation therefore is to rewrite the wave equation in an equivalent form where a scalar Laplacian \(\Delta\) is extracted from the two covariant derivatives, followed by additional coupling terms:
\begin{multline}\label{eq: general wave equation - component form}
\left(\Delta+k_{0}^{2}n_{0}^{2}\right)A^i
-\frac{1}{\sqrt{g}}\partial_{j}\sqrt{g}g^{ik}\partial_{k}A^{j}\\
+g^{ik}\partial_{k}\frac{1}{\sqrt{g}}\partial_{j}\sqrt{g}A^{j} 
=-\mu_{0}J^i
\end{multline}
Then, we follow \cite{Constrained_Particle_1981} and introduce a local set of coordinates to describe an immediate surroundings of an arbitrary surface -  \(\boldsymbol{R}\left(q_{1},q_{2},q_{3}\right)=\boldsymbol{r}\left(q_{1},q_{2}\right)+q_{3}\boldsymbol{\hat{n}}\left(q_{1},q_{2}\right)\), where the surface is parameterized by two coordinates \(\boldsymbol{r}=\boldsymbol{r}\left(q_{1},q_{2}\right)\) and \(\boldsymbol{\hat{n}}\) is the unit normal to the surface at a given point (see schematics in Figure \ref{fig: schematics}). In this coordinate system, the coefficients of the metric are found to be \cite{Constrained_Particle_1981}:
\begin{equation}\label{eq: 3d metric}
g_{ab}=\gamma_{ab}-2h_{ab}q_{3}+h_{ac}h_{b}^{\;c}q_{3}^{2},\hspace{0.5cm} g_{a3}=0, \hspace{0.5cm}g_{33}=1
\end{equation}
where \(a,b,c\) take the values of \(1,2\). \(\gamma_{ab}\) and \(h_{ab}\) are the components of the two-dimensional intrinsic metric and second fundamental form, respectively. The determinant of the metric is \(\sqrt{g}=\Omega\sqrt{\gamma}\), where \(\Omega=1-2Hq_3+K^2q_3^2\) \cite{Constrained_Particle_1981} is defined using the intrinsic \(K\) and extrinsic \(H\) curvatures.

Next, we assume that the surface has a dialectic waveguide attached to it, with a constant index \(n_0\), confining the light to travel within.
%Next, we assume that there is a curved surface dielectic waveguide following the surface defined by $g_{ab}$, with a constant index \(n_0\), confining the light to travel within it.
Finally, we evaluate Eq. \ref{eq: general wave equation - component form} using the metric defined by Eq. \ref{eq: 3d metric}. We take the narrow layer limit \(q_3\rightarrow0\) and arrive at the wave equations for the vector potential components (see Supplementary Material: S1). We separate the equations to the normal component:
\begin{equation}\label{eq: w.e normal component}
\left[\Delta_{\gamma}+\partial_{3}^{2}+K-3H^{2}+k_{0}^{2}n_{0}^{2}\right]\mathcal{A}^{3}-2\left(\partial_{a}H\right)\mathcal{A}^{a}=-\mu_{0}J^{3}
\end{equation}
and the two tangential components:
\begin{multline}\label{eq: w.e tangent components}
\left[\Delta_{\gamma}+\partial_{3}^{2}+H^{2}-K+k_{0}^{2}n_{0}^{2}\right]\mathcal{A}^{a}\\
+\left[\gamma^{ab}\partial_{b}\frac{1}{\sqrt{\gamma}}\partial_{c}\sqrt{\gamma}-\frac{1}{\sqrt{\gamma}}\partial_{c}\sqrt{\gamma}\gamma^{ab}\partial_{b}\right]\mathcal{A}^{c}\\
+2\left[\gamma^{ab}\left(\partial_{b}H\right)-h^{ab}\partial_{b}\right]\mathcal{A}^{3}
=-\mu_{0}J^{a}
\end{multline}
where \(\Delta_\gamma\) is the scalar Laplacian (or Laplace-Beltrami operator) with respect to the intrinsic metric. The field \(\mathcal{A}^i=\sqrt{\Omega}A^i\) was substituted in the wave equations to get the right volume measure after separating the normal component of the potential field \cite{Constrained_Particle_1981}.
The resulting equations couple the different $A^i$s, as expected for arbitrary coordinates. However, the coupling of the normal component is given only by \(\partial_aH\), hence, considering surfaces with constant extrinsic curvature eliminates the coupling. Therefore, by considering emitters' dipole pointing solely to the normal direction, the other components do not come into play, and equation \ref{eq: w.e normal component}'s Green function is enough to describe the interactions between all emitters. In \cite{Batz_Peschel_2008} it is shown that for the electric field, considering surfaces of revolution allows the decoupling of the azimuthal component in a similar manner, but we shall focus on the normal component to allow more general surfaces to be considered.

\begin{figure}
    \centering
    \includegraphics[width=0.8\linewidth]{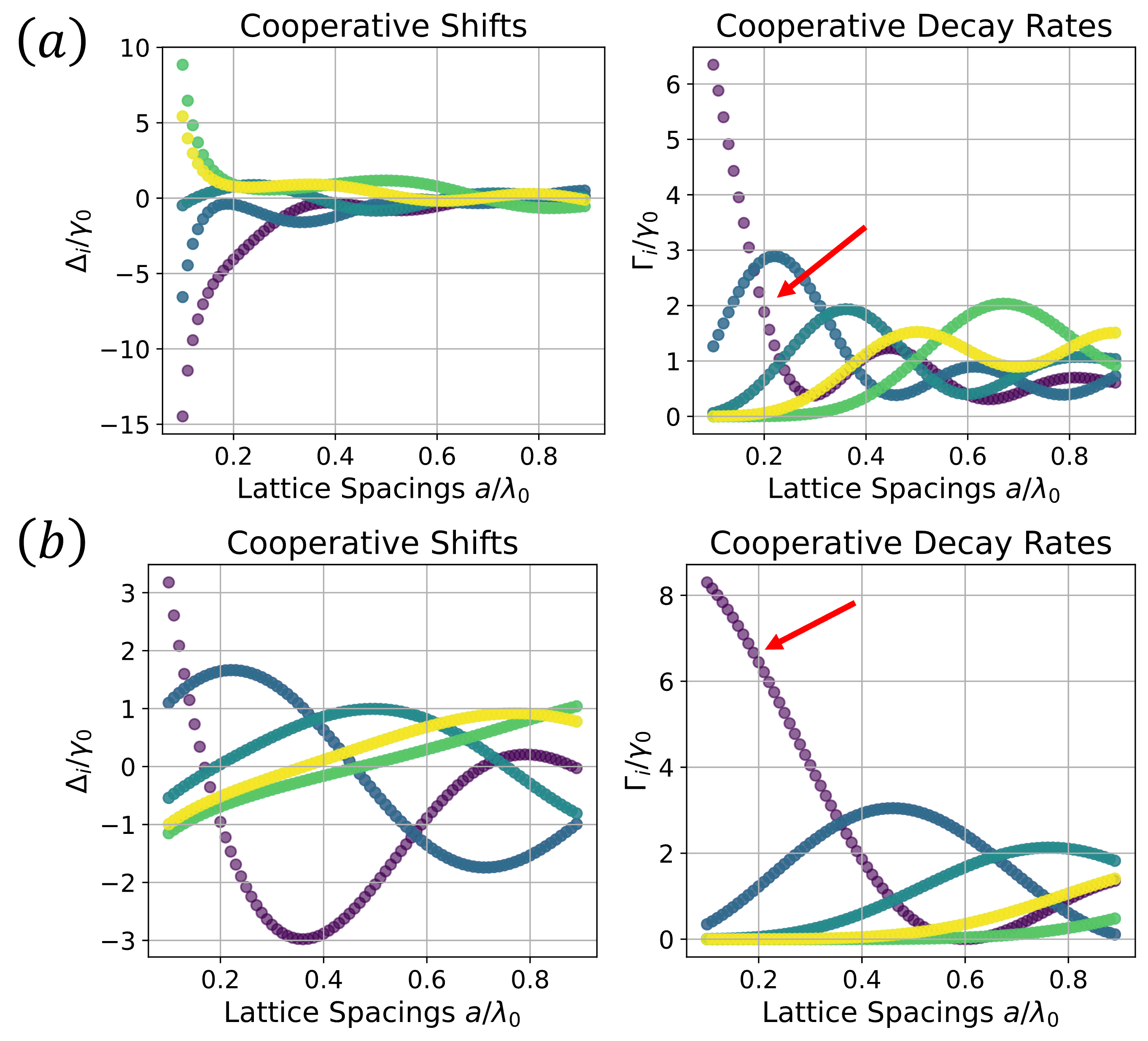}
    \caption{(a) Collective eigenvalues of a ring of emitters in free space. The emitters all are oriented to the \(\hat{z}\) direction. Left panel: Real part of the eigenvalues, corresponding to the collective shift of each excited state. Right panel: Imaginary part of the eigenvalues, corresponding to the collective decay rate of each state. At lattice spacing of \(a=0.2\lambda\), there are two superradiant states, the higher of which reaches \(\Gamma\approx3\gamma_{\text{free}}\). (b) Left panel: collective eigenvalues of the same ring of emitters embedded in a planar waveguide. the collective shifts of the ring are shown to have lower values compared to (a). In the right panel the collective decay rates are plotted, showing enhancement of the superradiant state. At \(a=0.2\), the higher superradiant state reaches \(\Gamma\approx6\gamma_{2d}\), a significant enhancement compared to the free space settings. Colors correspond to different eigenmodes.}
    \label{fig: freespace_vs_plane}
\end{figure}

\begin{figure}
    \centering
    \includegraphics[width=1\linewidth]{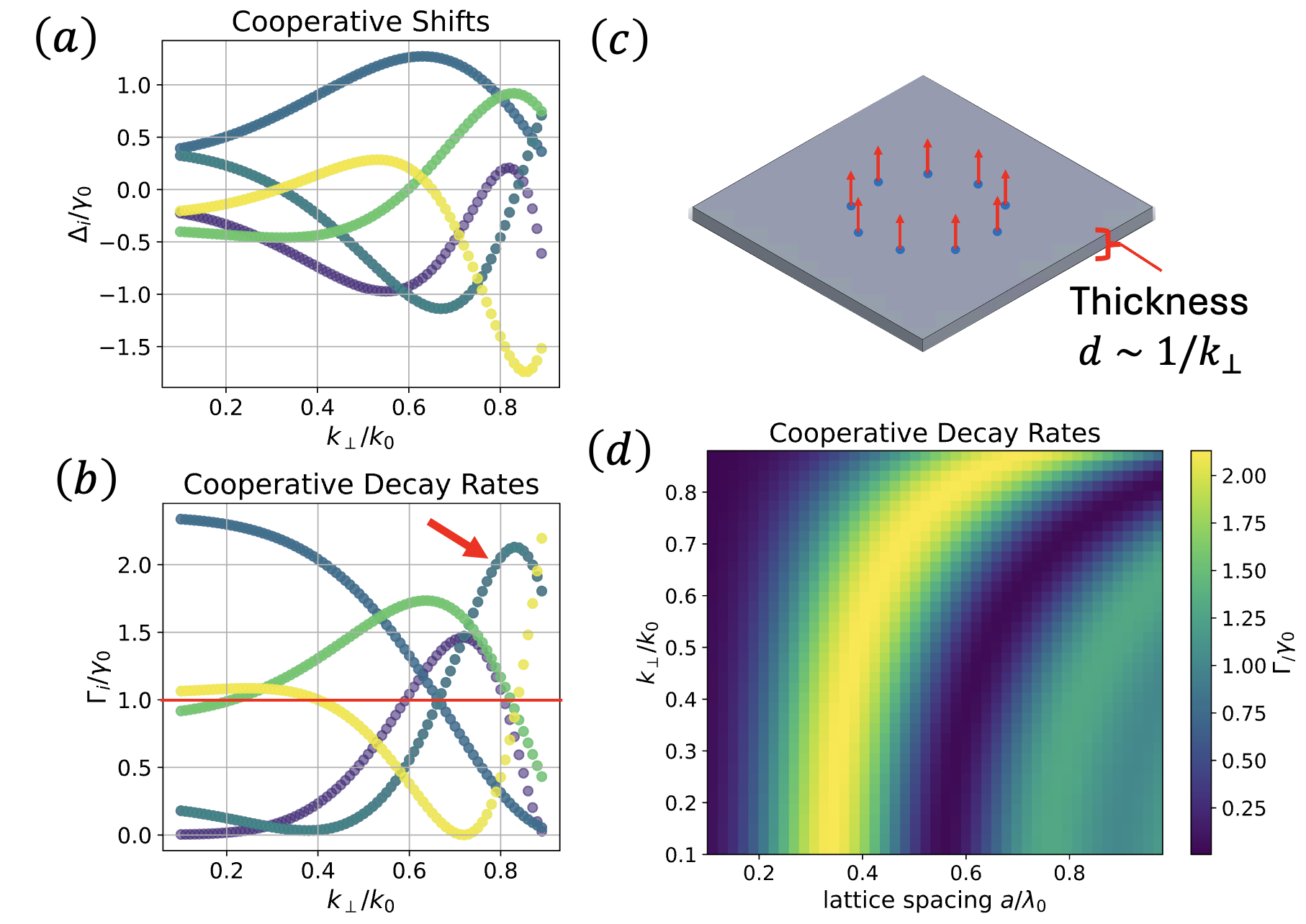}
    \caption{(a-b) Real and imaginary parts of collective eigenvalues for a ring of emitters on a planar waveguide, plotted versus normalized perpendicular momentum \(k_\perp / k_0\). The red line in (b) at \(\Gamma = \gamma_\perp\) marks the transition between subradiant and superradiant states, showing that varying \(k_\perp\) changes the nature of the state. The red arrow points at the state that is further analyzed in (d). (c) Schematics of the planar waveguide, where the thickness of the waveguide is shown. (d) Heat map of one eigenstate of the array, showing the decay rate of the state as a function of lattice spacing (x-axis) and \(k_\perp\) (y-axis). The map shows regions where the state is subradiant (blue) and a region where it is superradiant (yellow). Ring lattice spacing is \(a = 0.6\lambda_0\).}
    \label{fig: plane_k_perp}
\end{figure}

Assuming the extrinsic and intrinsic curvatures are relatively constant, substituting the separation \(\mathcal{A}^3\left(q_1,q_2,q_3\right)=\mathcal{A}\left(q_1,q_2\right)\mathcal{N}\left(q_3\right)\) in equation \ref{eq: w.e normal component}, and assuming one dominant guided surface mode (see Supplementary Material: S1 \cite{supp}), we finally obtain the wave equation of the normal component of the vector potential on a surface:
\begin{equation}\label{eq: simplified wave equation}
\left[\Delta_{\gamma}+k_{eff}^2\right]\mathcal{A}=-\mu_{0}J^{3}
\end{equation}
where the effective wavenumber is:
\begin{equation}\label{eq: effective wavenumber}
k_{eff}^2=k_{0}^{2}n_{0}^{2}-k_\perp^2+K-3H^{2}
\end{equation}
and the perpendicular momentum \(k_\perp\) is the constant of separation defined by \(\partial_3^2\mathcal{N}/\mathcal{N}=-k_\perp^2\). Equations \ref{eq: simplified wave equation} and \ref{eq: effective wavenumber} are main results of this paper. They show how the curvatures of a surface affect the effective wavelength of the normal component of the potential field. Once a Green function to equation \ref{eq: simplified wave equation} is found (either in summation representation or a closed-form expression), the normal component of the dyadic Green function is easily obtained (see \cite{supp} S2), and it can be directly substituted in equation \ref{eq: effective Hamiltonian} to obtain the cooperative eigenvalues of a collection of emitters.
Notably, the curvature term \(K-3H^2\) in the effective wavelength is different than the corresponding term found for the electric field wave equation \(H^2\) \cite{Batz_Peschel_2008}.
As noted in \cite{schultheiss2020light} for the electric field, the curvatures contribute to the effective wavelength only when they are on the same scale as \(k_0\). Otherwise, \(k_0\) dominates, and the intrinsic curvature will only affect the wave equation through the Laplace-Beltrami operator \(\Delta_\gamma\).\\

%\section{\label{sec: planar waveguide}Cooperative Effects in Planar Waveguides}
As a first example, we consider the simplest surface imaginable to demonstrate the significance of the perpendicular momentum \(k_\perp\) when calculating cooperative effects on a waveguide. The exact value of this momenta component will be dictated by the actual layer structure of the waveguide and the relevant refractive indexes, but we assume a single-mode planar waveguide for which this momentum component is inversely proportional to the thickness of the waveguide \(k_\perp\sim 1/d\). The assumption of a single-mode waveguide is consistent with the narrow-layer limit \(q_3\rightarrow0\).
The Green function of equation \ref{eq: simplified wave equation} when considering a plane is known to be \cite{linton1998green}:
\begin{equation}\label{eq: plane_green_function}
G_{\text{plane}}\left(\boldsymbol{r},\boldsymbol{r}^\prime\right)=\frac{i}{4}H_{0}^{\left(1\right)}\left(k_\text{eff}r\right)
\end{equation}
where \(H^{\left(1\right)}_0\) is the Hankel function of the first kind, \(r\equiv\left|\boldsymbol{r}-\boldsymbol{r}^\prime\right|\) is the distance from the source and \(k_\text{eff}\) is the same as in equation \ref{eq: effective wavenumber} , with \(K=H=0\). The dyadic Green function for the plane is obtained from \ref{eq: plane_green_function} (\cite{supp} S2), and then substituted in the effective Hamiltonian \ref{eq: effective Hamiltonian}. This Hamiltonian is diagonalized, and the complex eigenvalues are interpreted as the cooperative shifts (real part) and cooperative decay rates (imaginary part) of the eigenstates. We distinguish two types of states: states with high decay rates $>1$ (when normalized by the bare atom decay) are considered supperradiant, and states with low decay rate  $<1$ are considered subradiant. In figure \ref{fig: freespace_vs_plane} we plot the cooperative eigenvalues of a ring of emitters, for free space and for plane waveguide with large perpendicular momentum \(k_\perp=0.9k_0n_0\), as a function of the separation \(a\) between the atoms. In panels b and d, the decay rates of the states are shown, and red arrows point to the most supperadiant state of the array in each case. Comparing this most radiant state, one can see that the presence of the waveguide enhances its decay rate significantly (by about 3-times for \(a=0.2\lambda_0\)). The waveguide also changes the functional dependence of some of the cooperative shifts on the lattice spacing. In figure \ref{fig: plane_k_perp}, we display the collective eigenvalues of the same ring for constant spacing, for different perpendicular momentum \(k_\perp\). One can clearly see that varying \(k_\perp\) greatly affects the collective eigenvalues, and can even make subradiant states into superradiant and vice-versa. This significance carry on to any surface, not necessarily a plane. \\

\begin{figure}
    \centering
    \includegraphics[width=1\linewidth]{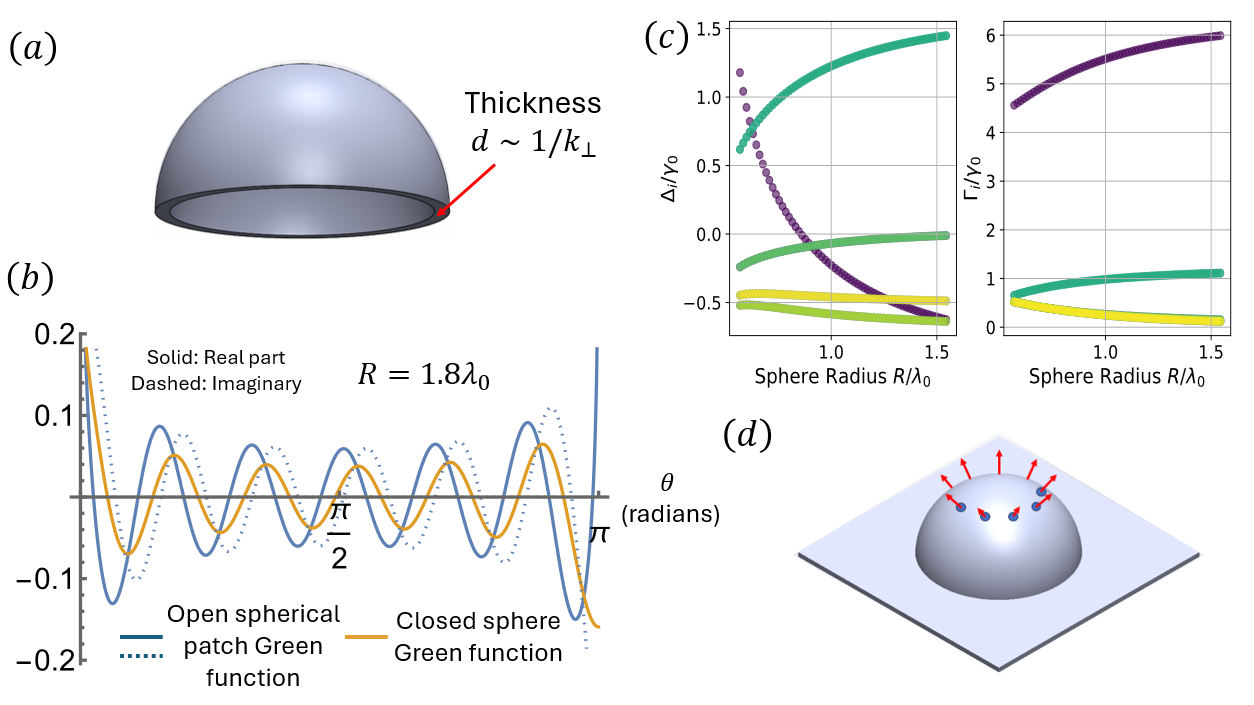}
    \caption{(a) Sketch of a locally spherical waveguide, where the thickness of the waveguide is shown. (b) Comparison between the known Helmholtz Green function of the sphere (yellow) to our new propagating Green function (blue). The closed surface Green function does not diverge on the opposite pole, but has no imaginary component, making it a standing wave solution. The locally spherical but globally open Green function is a propagating solution, but diverges on the opposite pole. This shows that this Green function can not be used for large central angles. This assumption is consistent with the requirement that the atoms are closely spaced.
(c) Real and imaginary parts of eigenvalues of the collective eigenvalues of a ring of emitters on a sphere as a function of sphere radius \(R\), plotted for \(k_\perp=0.9k_0\), corresponding to a thin waveguide. The lattice constant \(a\), which is the arclength between neighboring emitters is kept fixed. At high curvatures the collective decay rates are suppressed, as the superradiant states are less radiant and the subradiant states are less long lived. (d) sketch of the same locally spherical waveguide, showing the transition from curved to open surface, required for outwards propagation.}
    \label{fig: curved surfaces}
\end{figure}
%\section{\label{sec: positive curvature waveguide}Waveguide with Constant Positive Curvature}
We shall now demonstrate how our formalism enables new Green functions to be found for curved surfaces. The most natural surface to consider is the sphere, as it has constant intrinsic curvature \(K=1/R^2\)  and constant extrinsic curvature \(H=1/R\), where \(R\) is the radius of the sphere. Moreover, as the sphere is homogeneous and isotropic, we can expect to find a closed form Green function, without needing to resort to summations. Using the spherical coordinates \(\theta,\phi\), equation \ref{eq: simplified wave equation} becomes the Legendre equation, with the Legendre functions of the first kind \(P_\nu\) and second kind \(Q_\nu\) as fundamental solutions. The degree \(\nu\) of the Legendre functions is given by \(\nu\left(\nu+1\right)=k_{eff}^2R^2\), and the effective wavelength has a contribution from the curvatures \(k_{eff}^2=k_{0}^{2}n_{0}^{2}-k_\perp^2-2/R^{2}\). The closed sphere has a known Helmholtz Green function, given by \(G_{\text{sphere}}\left(\boldsymbol{r},\boldsymbol{r}^\prime\right)=\frac{1}{2\pi}P_\nu\left(-\cos\theta\right)\), where \(\cos\theta=\boldsymbol{r}\cdot\boldsymbol{r}^\prime/R^2\) is the cosine of the central angle between the two points. The Legendre function of the second kind \(Q_\nu\) is discarded as it diverges on the antipodal point of the point current. One can immediately see however that this Green function does not describe a radiating solution, as it is a real function, thus describing a standing wave solution. Indeed, on a closed surface, there is no such notion of a radiating solution as there is nowhere for the field to propagate out to. To deal with this we consider a region of a surface which is locally a sphere, but globally unbounded (see figure \ref{fig: curved surfaces}a). In such a region, we assume the Legendre equation to hold locally, and span our Green function on both kinds of the Legendre functions. At this point one would normally use the Sommerfeld radiation condition to ensure outward propagating solutions. We propose here a different method to determine the coefficients, where we assume the coefficients to be independent of the radius of curvature, and demand that in the limit of large radii, the Green function will reproduce the known Green function of the plane (equation \ref{eq: plane_green_function}). Formally, we employ known limits of the Legendre functions, transforming them into Bessel functions (see \cite{abramowitz2012handbook}). Using these limits, we can correctly deduce the coefficients of a radiating solution on a locally spherical surface patch:
%We now consider a region of a surface with positive constant positive curvature, making it locally a sphere. The surface is assumed to be globally unbounded to allow radiating solutions. For such a surface patch, equation \ref{eq: w.e normal with k effective} becomes the Legendre equation, with solutions spanned over the Legendre functions of the first kind \(P_\nu\) and second kind \(Q_\nu\) . To determine the coefficients of the radiating Green function, we take the limit of small curvatures, and demand that the Green function will reproduce equation \ref{plane_green_function}. Using known identities of the Legendre functions \cite{abramowitz2012handbook}, we find the radiating Green function of a spherical surface patch to be:
\begin{equation}
G_\text{sphere}\left(\theta\right)=\frac{1}{2\pi}Q_{\nu}\left(\cos\theta\right)+\frac{i}{4}P_{\nu}\left(\cos\theta\right)
\end{equation}
where as before \(\nu\left(\nu+1\right)=k_{eff}^2R^2\). To the best of our knowledge, This is the first presentation of a radiating Green function on a locally spherical surface patch. In figure 4b we display a comparison between both Green functions of a sphere. The complex-valued open spherical Green function is shown to diverge on the opposite pole, while the real-valued closed Green function does not. In \ref{fig: curved surfaces}c, we display the calculated collective eigenvalues for the ring of emitters embedded on a sphere as a function of increasing sphere radii, while the arclength \(a\) between emitters is kept fixed. As expected, the curvature of the sphere affects the cooperative eigenvalues only when it is comparable to the wavelength. However, the curvature cannot be taken to be arbitrarily large. Eventually we require the surface to “flatten out” to allow open boundary conditions, introducing derivatives of the curvature terms and breaking the assumptions of constant curvatures (see figure 4a). For a small number of very closely spaced emitters (\(\sim0.1\lambda_{0}\)), it is feasible to use large curvatures, as done in figure 4c. 

We stress here that the sphere was chosen for demonstration because it allowed us to present a closed-form Green function. This method will work for all surfaces that fulfill our constant curvature assumptions and open boundary conditions, but one might have to resort to summation methods to represent the Green function instead of a closed-form expression. The main advantage of the present formalism applies to all such surfaces - it replaces the challenge of determining the dyadic Green function for three coupled Helmholtz equations with the simpler task of finding a scalar Green function for a single Helmholtz equation.\\

We now discuss the practicality of our analysis to experiments. Curved waveguides that fits our formalism and assumptions have been demonstrated in experiments to support perpendicular momentum of up to \(k_\perp\sim0.6k_0\) \cite{Bekenstein_Curved_Space_2017, roth2024non}. A potential implementation system are quantum dot arrays that can be grown  by their self-assembly feature in thin and curved films \cite{zhang2022large}. The requirement that the dipole matrix elements will predominantly point to the normal of the surface is an experimental challenge. For quantum dots one can employ the less energetic dipole-allowed transition \cite{becker2018bright}. Another potential emitters are defects in crystals, for example silicon vacancies color centers in diamond \cite{RN89}, that are known to have definite dipole orientations within the lattice. Photon-mediated interactions between silicon vacancy color centers was established in nanophotonic cavities \cite{RN7} and in a deflectable nanophotonic waveguide \cite{machielse2019quantum} that can be interpreted as curved waveguides. The inclusion of both curvature control and definite dipole orientations make quantum dots and defects in crystals potential platforms for experimental implementation of our work.\\ 
%\section{\label{conclusions}Conclusions}
In conclusion, we have derived the vector potential wave equation constrained to an arbitrary surface, and showed that the normal component is decoupled under the assumption of constant extrinsic curvature. %By requiring all dipoles to be oriented to the normal direction, we have identified the contributing dyadic Green function component and the effective wavelength. 
We identified the contributing component of dyadic Green function for normal dipoles orientation and the effective wavelength of the photon interaction. Our work shows that the perpendicular momentum \(k_\perp\) have great significance over the atomic collective states, both in enhancing superradiant states and in dictating the nature of the collective state (transforming superradiant states into subradiant states and vice versa), suggesting a method to control the collective states nature.  We have analyzed how large curvature settings affect the photon-mediated interactions by deriving a novel Green function for a locally spherical surface patch. A combination of boundary conditions and curvature settings should be further explored for control over the effective wavelength, including negative Gaussian curvature. State-of-the-art systems aim to embed emitters or atomic arrays in nanophotonics architectures \cite{nano_photonic_integration_0_Hannes_integrated_2024} that require high curvatures, for specific applications  \cite{zhang2022large}. The formalism developed in this work is useful to model emitter arrays embedded in arbitrary two-dimensional nanostructures with high curvature and is expected to have great significance in designing and predicting the optical response of systems that integrate atomic arrays and nanophotonics structures \cite{nano_photonic_integration_0_Hannes_integrated_2024}. 
%
%For example this can be implemented by embedding emitters in thin curved films \cite{Bekenstein_Curved_Space_2017}. %which are known for their ability to self-organize into arrays \cite{raino2018superfluorescence}.
% If in two-column mode, this environment will change to single-column
% format so that long equations can be displayed. Use
%\begin{widetext}
%\end{widetext}

% figures should be put into the text as floats.
% Use the graphics or graphicx packages (distributed with LaTeX2e)
% and the \includegraphics macro defined in those packages.
% See the LaTeX Graphics Companion by Michel Goosens, Sebastian Rahtz,
% entire page. There is no need to do explicit centering.

% \begin{figure}
% \end{figure}

% Surround figure environment with turnpage environment for landscape
% figure
% \begin{figure}
% \end{figure}

%\appendix
%\section{}

%\begin{acknowledgments}
%\end{acknowledgments}

\bibliography{Main.bib}

\end{document}